\documentstyle[epsfig]{mn}

\newif\ifAMStwofonts

\def\gtsima{$\; \buildrel > \over \sim \;$}
\def\ltsima{$\; \buildrel < \over \sim \;$}
\def\gsim{\lower.5ex\hbox{\gtsima}}
\def\lsim{\lower.5ex\hbox{\ltsima}}
\newcommand{\etal}{et al.\ }
\def\Msun{{M_\odot}}
\def\Zsun{{Z_\odot}}
\def\be{\begin{equation}}
\def\ee{\end{equation}}
\def\ie{{\frenchspacing\it i.e. }}

\ifoldfss
  \ifCUPmtlplainloaded \else
    \NewTextAlphabet{textbfit} {cmbxti10} {}
    \NewTextAlphabet{textbfss} {cmssbx10} {}
    \NewMathAlphabet{mathbfit} {cmbxti10} {} 
    \NewMathAlphabet{mathbfss} {cmssbx10} {} 
  \fi
  \ifAMStwofonts
    \ifCUPmtlplainloaded \else
      \NewSymbolFont{upmath} {eurm10}
      \NewSymbolFont{AMSa} {msam10}
      \NewMathSymbol{\upi}     {0}{upmath}{19}
      \NewMathSymbol{\umu}     {0}{upmath}{16}
      \NewMathSymbol{\upartial}{0}{upmath}{40}
      \NewMathSymbol{\leqslant}{3}{AMSa}{36}
      \NewMathSymbol{\geqslant}{3}{AMSa}{3E}

      \let\geq=\geqslant \let\ge=\geqslant
    \fi
  \fi
\fi 

\ifnfssone
  \newmathalphabet{\mathit}
  \addtoversion{normal}{\mathit}{cmr}{m}{it}
  \addtoversion{bold}{\mathit}{cmr}{bx}{it}
  \newmathalphabet{\mathbfit} 
  \addtoversion{normal}{\mathbfit}{cmr}{bx}{it}
  \addtoversion{bold}{\mathbfit}{cmr}{bx}{it}
  \newmathalphabet{\mathbfss} 
  \addtoversion{normal}{\mathbfss}{cmss}{bx}{n}
  \addtoversion{bold}{\mathbfss}{cmss}{bx}{n}
  \ifAMStwofonts
    \ifCUPmtlplainloaded \else
      \UseAMStwoboldmath
      \makeatletter
      \new@mathgroup\upmath@group
      \define@mathgroup\mv@normal\upmath@group{eur}{m}{n}
      \define@mathgroup\mv@bold\upmath@group{eur}{b}{n}
      \edef\UPM{\hexnumber\upmath@group}
      \new@mathgroup\amsa@group
      \define@mathgroup\mv@normal\amsa@group{msa}{m}{n}
      \define@mathgroup\mv@bold\amsa@group{msa}{m}{n}
      \edef\AMSa{\hexnumber\amsa@group}
      \makeatother
      \mathchardef\upi="0\UPM19
      \mathchardef\umu="0\UPM16
      \mathchardef\upartial="0\UPM40
      \mathchardef\leqslant="3\AMSa36
      \mathchardef\geqslant="3\AMSa3E

      \let\geq=\geqslant \let\ge=\geqslant
    \fi
  \fi
\fi 

\ifnfsstwo
  \DeclareMathAlphabet{\mathbfit}{OT1}{cmr}{bx}{it}
  \SetMathAlphabet\mathbfit{bold}{OT1}{cmr}{bx}{it}
  \DeclareMathAlphabet{\mathbfss}{OT1}{cmss}{bx}{n}
  \SetMathAlphabet\mathbfss{bold}{OT1}{cmss}{bx}{n}
  \ifAMStwofonts
    \ifCUPmtlplainloaded \else
      \DeclareSymbolFont{UPM}{U}{eur}{m}{n}
      \SetSymbolFont{UPM}{bold}{U}{eur}{b}{n}
      \DeclareSymbolFont{AMSa}{U}{msa}{m}{n}
      \DeclareMathSymbol{\upi}{0}{UPM}{"19}
      \DeclareMathSymbol{\umu}{0}{UPM}{"16}
      \DeclareMathSymbol{\upartial}{0}{UPM}{"40}
      \DeclareMathSymbol{\leqslant}{3}{AMSa}{"36}
      \DeclareMathSymbol{\geqslant}{3}{AMSa}{"3E}

      \let\geq=\geqslant \let\ge=\geqslant
    \fi
  \fi
\fi 

\ifCUPmtlplainloaded \else
  \ifAMStwofonts \else 
    \def\upi{\pi}
    \def\umu{\mu}
    \def\upartial{\partial}
  \fi
\fi

\title{Ultra faint dwarfs: probing early cosmic star formation}
\author[Stefania Salvadori \& Andrea Ferrara]
{Stefania Salvadori$^{1}$ \& Andrea Ferrara$^{2,3}$\\
$^1$SISSA/International School for Advanced Studies, Via Beirut 4, 34100 Trieste, Italy\\ 
$^2$Scuola Normale Superiore, Piazza dei Cavalieri 7, 56126 Pisa, Italy\\
$^3$Blaauw Professor, Kapteyn Astronomical Institute, Groningen, The Nederlands}
\date{}

\pagerange{\pageref{firstpage}--\pageref{lastpage}}
\pubyear{}

\begin{document}

\maketitle 
\label{firstpage}

\begin{abstract}
We investigate the nature of the newly discovered Ultra Faint dwarf spheroidal 
galaxies (UF dSphs) in a general cosmological context simultaneously accounting 
for various ``classical`` dSphs and Milky Way properties including their
Metallicity Distribution Function (MDF). To this aim we extend the merger tree 
approach previously developed to include the presence of star-forming minihaloes, 
and a heuristic prescription for radiative feedback. The model successfully 
reproduces both the observed [Fe/H]-Luminosity relation and the mean MDF of UFs. 
In this picture UFs are the oldest, most dark matter-dominated ($M/L > 100$) dSphs 
with a total mass $M= 10^{7-8}\Msun$; they are leftovers of H$_2$-cooling 
minihaloes formed at $z > 8.5$, i.e. before reionization. Their MDF is broader 
(because of a more prolonged SF) and shifted towards lower [Fe/H] (as a result of 
a lower gas metallicity at the time of formation) than that of classical dSphs. 
These systems are very ineffectively star-forming, turning into stars by $z=0$ 
only $<3\%$ of the potentially available baryons. We provide a useful fit for the 
star formation efficiency of dSphs.
\end{abstract}


\section{Background}
The dwarf spheroidal galaxy (dSph) population observed in the Milky Way (MW) system 
provides crucial insights into cosmic structure formation. Nevertheless, the origin 
and evolution of dSphs has so far escaped any comprehensive theoretical 
interpretation and the picture is now further complicated by the discovery of a 
new class of dwarf satellite galaxies: the Ultra Faint dSphs (UFs).

UFs are the least luminous galaxies known, with a total luminosity 
$L\approx 10^{3-5} L_{\odot}$; spectroscopic follow-up has revealed that they are 
highly dark matter dominated systems $M/L > 100$ (Simon \& Geha 2007, Geha et~al. 2008). 
Their average iron-abundance is $\langle$[Fe/H]$\rangle <-2$ (Kirby et~al. 2008) 
\ie they represent the most metal-poor stellar systems ever known; although more data 
are required to solidly constrain stellar populations, at the moment {\it all} of them
appear to be dominated by an old stellar population (Walsh, Willman \& Jerjen 2008),
with the only exception of LeoT (de Jong et~al. 2008).
In addition, UFs are relatively common in the MW system, representing more than $50\%$ of the total 
number of dSph companions discovered so far ($N\approx 20$); hence, they are not peculiar objects. 
When and how did UFs form? 

The [Fe/H]-Luminosity relation derived for UFs (Kirby et~al. 2008) constitutes an extension towards 
lower metallicity of that of ``classical`` dSphs ($L >10^5L_{\odot}$). Such a continuous trend seems 
to exclude that processes (e.g. tidal stripping) different from those shaping the relation for 
classical dSphs become dominant in these objects. However, while dSphs and UFs together span more 
than four orders of magnitude in luminosity, their total mass is roughly the same $M\approx 10^7\Msun$ 
within the innermost 300 pc (Strigari et~al. 2008, Li et~al. 2008). Why is then star formation so 
inefficient in UF satellites? 

The puzzle is made even more intriguing by the observation of metal-poor stars in UFs 
(Kirby et~al. 2008) which have revealed the existence of a [Fe/H]$<-3$ stellar population, 
in contrast with the classical dSphs which are lacking extremely iron-poor stars (Helmi et~al. 2006). 
Does this reflect a different origin of UF and classical dSphs?
In this Letter we extend previous works (Salvadori, Schneider, Ferrara 2007 [SSF07]; Salvadori,
Ferrara, Schneider 2008 [SFS08]) to investigate the origin of UF galaxies in a cosmological context.

\section{Model Features}
In this Section we will briefly describe the basic features of the model implemented in the code GAMETE 
(GAlaxy MErger Tree \& Evolution) which allows us to reconstruct the stellar population history and the 
chemical enrichment of the MW along its hierarchical merger tree, following at the same time the 
formation and evolution of dSph satellites\footnote{We adopt a $\Lambda$CDM cosmological model with 
$h=0.73$, $\Omega_{m}=0.24$, $\Omega_{\Lambda} = 0.72$, $\Omega_{b}h^{2}=0.02$, $n=0.95$ and 
$\sigma_8 = 0.74$, consistent with WMAP5 data (Spergel \etal 2006); we assume the fit to the power spectrum 
proposed by Bardeen \etal (1986) and modified by Sugiyama (1995).}. In such a model dSphs form 
out of the MW environment, or Galactic Medium (GM), whose metal enrichment strongly depends on the star 
formation (SF) history and mechanical feedback along the hierarchical tree. The model successfully 
reproduces the global properties of the MW and the Metallicity Distribution Function (MDF) of Galactic 
halo stars, along with the observational properties of prototype dSphs, as for example Sculptor (SFS08).

The merger history of the MW is reconstructed starting from $z=20$ 
via a Monte Carlo algorithm based on the extended Press-Schechter theory and 
by adopting a binary scheme with accretion mass (Volonteri, Haardt \& Madau 2003).
The evolution of gas and stars is then followed along the hierarchy with
the following hypotheses: (a) stars can only form in haloes of mass 
$M_h > M_{sf}(z)$; the evolution of $M_{sf}(z)$ will be discussed in the next Section;
(b) in each halo the SF rate is proportional to the mass of cold gas whose gradual accretion 
is described by a numerically calibrated infall rate (Kere\v{s} et al. 2005); 
(c) according to the so-called critical metallicity scenario (Schneider et~al. 2002, 2006)
low-mass stars with a Larson Initial Mass Function form when the gas metallicity 
$ Z \ge Z_{\rm cr}=10^{-5\pm 1}\Zsun$; below $Z_{\rm cr}$ massive Pop~III stars form with 
a reference mass $m_{\rm PopIII}=200\Msun$.
The chemical enrichment of the gas is followed by taking into account mass-dependent
stellar evolutionary timescales (Lanfranchi \& Matteucci 2007).
We follow the chemical evolution of gas both in proto-Galactic haloes and in the GM by including a 
physically based description of the mechanical feedback due to supernova (SN) energy deposition 
(see SFS08 for details). Metals and gas are assumed to be instantaneously and homogeneously 
mixed with the gas (implications discussed in SFS08). 
The code is calibrated by simultaneously reproducing the global properties of the MW 
(stellar/gas mass and metallicity) and the MDF of metal poor stars observed by Beers 
\& Christlieb (2006) in the Galactic halo (for details see SSF07).

\begin{figure}
  \centerline{\psfig{figure=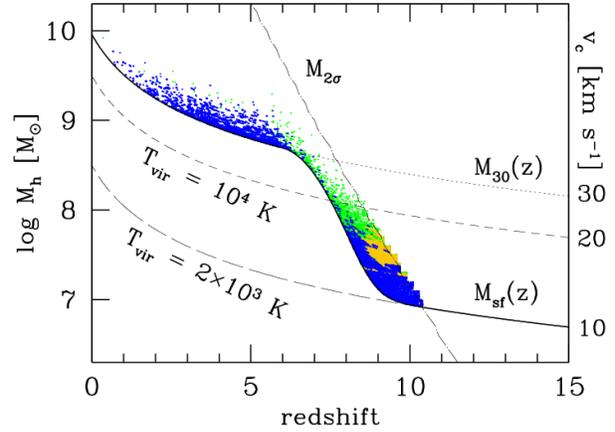,width=8cm,angle=0}}
  \caption{Dark matter halo mass and circular velocity of selected dSph candidates as a function of their 
    formation redshift $z$ (points) for 10 realizations of the hierarchical merger tree. Different colours 
    show the baryonic fraction $f_b$ at the formation epoch with respect to the cosmic value $f_c=0.156$: 
    $f_b/f_c > 0.5$ (blue), $0.1 < f_b/f_c < 0.5$ (green), $f_b/f_c < 0.1$ (yellow). The lines show the 
    evolution of $M_{sf}(z)$ (solid), $M_{30}(z)$ (dotted), the halo mass corresponding to 2$\sigma$ peaks 
    (dotted-long dashed), $T_{vir}=10^4$K (short dashed line) and $T_{vir}=2\times 10^3$K (long dashed line).} 
  \label{fig:1}
\end{figure}
\begin{figure}
  \centerline{\psfig{figure=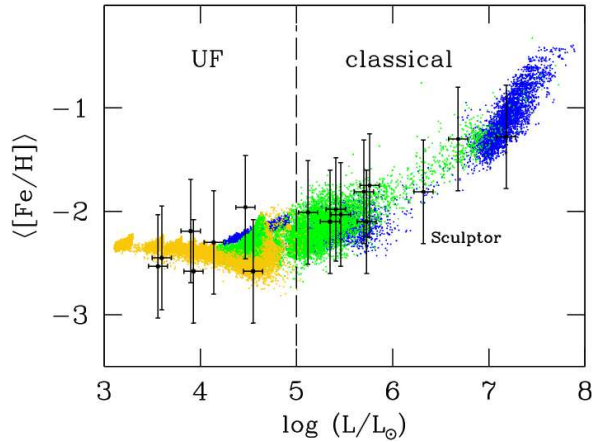,width=8.0cm,angle=0}}
  \caption{Total luminosity of selected dSph candidates as a function of their average iron abundance (points) 
    for 10 realizations of the hierarchical merger tree. As in Fig.~1 different colours show the 
    baryonic fraction at the formation epoch with respect to the cosmic value. The points with error bars are 
    observational data from Kirby et~al. 2008.} 
  \label{fig:2}
\end{figure}
Dwarf spheroidal candidates are selected among star forming haloes ($M_h >M_{sf}$) which are likely to 
become satellites \ie those corresponding to density fluctuations $<2\sigma$\footnote{The quantity 
  $\sigma(M,z)$ represents the linear rms density fluctuation smoothed with a top-hat filter of mass 
  $M$ at redshift $z$.}. Such a dynamical argument is based on N-body cosmological simulations by 
Diemand, Madau \& Moore (2005). Therefore, at each redshift, we select dSph candidates from the merger 
tree among haloes with $M_{sf}<M_h<M_{2\sigma}$; we then follow the isolated evolution (no further merging 
or accretion events) of ´´virtual¨ haloes with the same initial conditions (dark matter/gas/stellar 
content, metallicity). Through this method we build a statistically significant dSph sample; however, 
it prevents us from making specific predictions on the actual number of dSph satellites.  
\subsection{Feedback and star formation}
Radiative feedback processes are crucial to determine the minimum mass, $M_{sf}$, and efficiency, $f_*$, 
for star formation. The redshift evolution of the minimum mass of star-forming haloes, $M_{sf}$, is 
determined by two distinct radiative feedback processes. The first one has to do with the increase of 
the Jeans mass in progressively ionized cosmic regions; as a result, the infall of gas in haloes below a 
given circular velocity, $v^*_c$, is quenched and such objects are gas-starved. The evolution of 
$v^*_c(z)$ depends on the details of the reionization history (Gnedin 2000, Schneider et~al. 2008); 
we adopt the typical value $v^*_c=30$~kms$^{-1}$ after the end of reionization $z_{rei}=6$ \ie we assume 
$M_{sf}(z)=M_{30}(z)$ when $z<6$. Before reionization a second type of feedback, related to the 
photodissociation of H$_2$ molecules by the Lyman-Werner (LW) background photons, becomes important. 
As this species is the only available coolant for "minihaloes'' ($T_{vir}<10^4$~K) SF is suppressed in 
these objects to an extent which depends on the intensity of the UV background (Haiman, Rees \& Loeb 1996; 
Ciardi, Ferrara \& Abel 2000; Kitayama et~al. 2000; Machacek, Bryan \& Abel 2001). According to Dijkstra 
et~al. (2004) at $z\approx 10$ objects with $v_c\geq 10$~kms$^{-1}$ ($T_{vir}\approx 2000$K) can self-shield 
and collapse; therefore we use this value as the minimum absolute threshold for star formation. 
During reionization ($6<z<9$) the interplay between these two feedback types is quite complicated and 
no consensus is found on the evolution of $M_{sf}(z)$ (see for example Fig.~25 of Ciardi \& Ferrara 2005).
Instead of modeling in detail the build-up of LW and ionizing UV backgrounds, we interpolate between the 
low- and high-redshift behaviors and use an heuristic form for $M_{sf}(z)$ which leads to the suppression 
of SF in gradually more massive objects (Fig.~1). Such parameterization of $M_{sf}(z)$ allows to correctly 
reproduce the observed iron-luminosity relation of dSphs (Fig.~2)\footnote{The total luminosity value is 
derived from the stellar mass content as $L=M_*\times (M/L)_*$ by assuming $(M/L)_* = 1$.}. Fig.~1 also 
shows the halo masses of selected dSph candidates along with their initial baryonic fraction, $f_b$, with 
respect to the cosmic value $f_c=\Omega_b/\Omega_M=0.156$. We classify dSphs according to their initial 
baryonic content as gas-rich, $f_b/f_c > 0.5 $, intermediate, $0.1 < f_b/f_c < 0.5$, and gas-poor, 
$f_b/f_c < 0.1$, systems.
\begin{figure*}
  \centerline{\psfig{figure=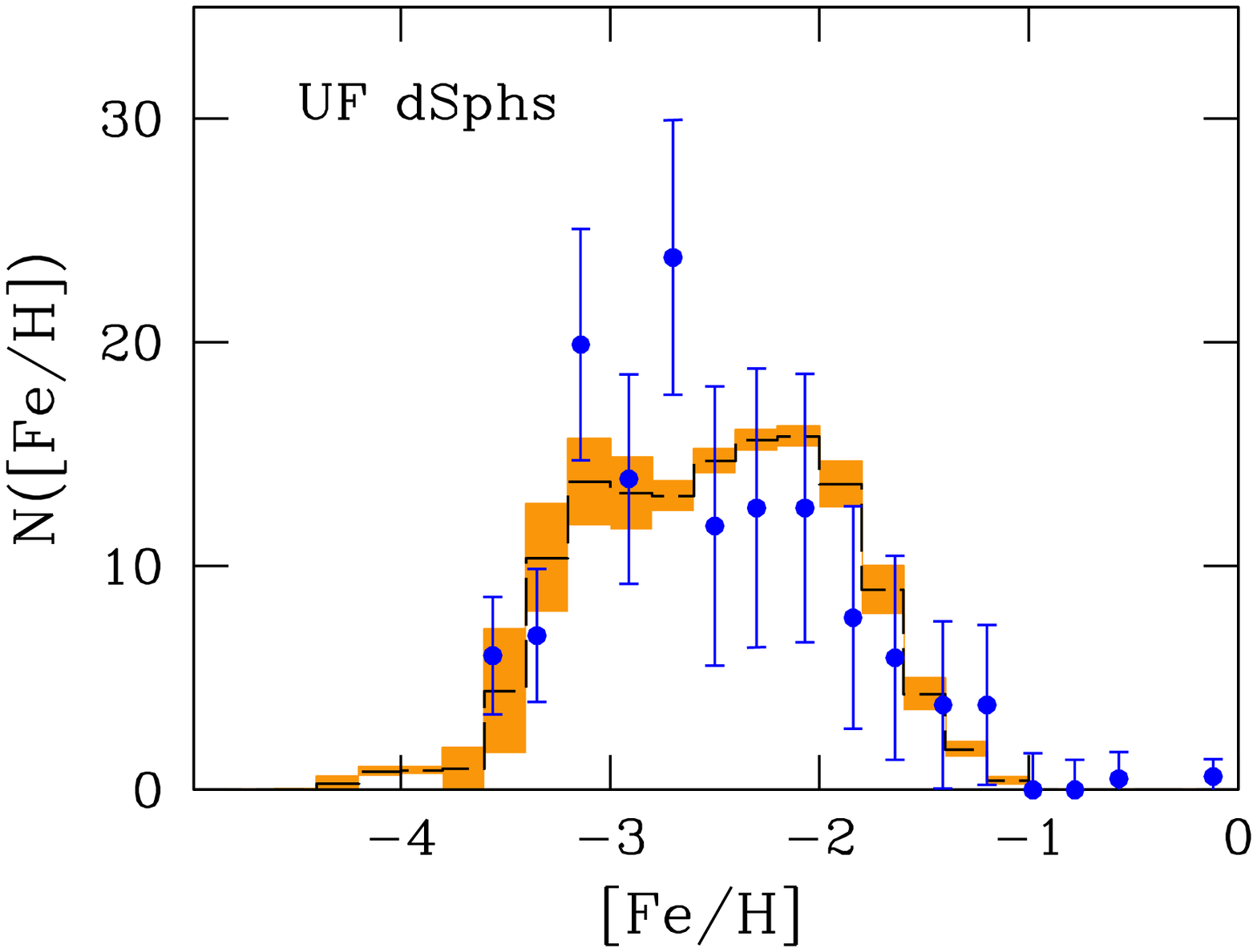,width=7.95cm,angle=0}
    \psfig{figure=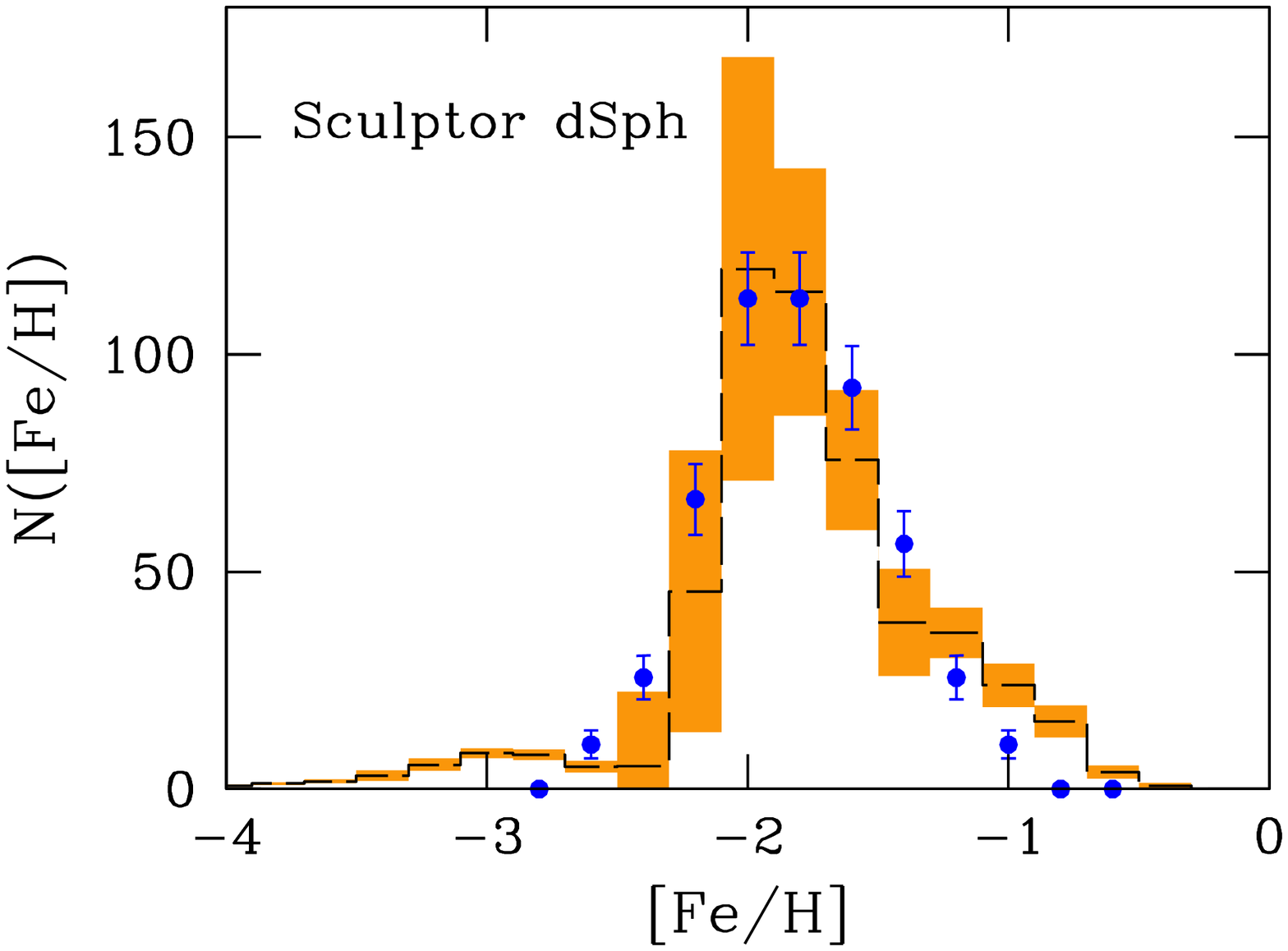,width=7.95cm,angle=0}}
  \caption{{\it Left panel:} comparison between the MDF of UFs observed by Kirby et~al. (2007) (points) 
    and the simulated one (histogram). Error bars are the Poissonian errors. The histogram is the averaged
    MDF over all UFs candidates ($L< 10^5 L_{\odot}$) in 10 realizations of the merger tree. The shaded 
    area represents the $1\sigma$ scatter among different realizations.
    {\it Right panel} comparison between the Sculptor MDF observed 
    by Helmi et~al. (2006) (points), and the simulated one (histogram). Error bars are the Poissonian 
    errors. The histogram is the averaged MDF over all the Sculptor-like dSph candidates 
    ($10^6 L_{\odot}<L<10^{6.5} L_{\odot}$) in 10 realizations of the merger tree. 
    The shaded area represents the $\pm 1\sigma$ Poissonian error.}
\label{fig:3}
\end{figure*} 
A second important feature of the model is that we consider a mass-dependent SF efficiency. 
In minihalos, in fact, the ineffective cooling by H$_2$ molecules limits the 
amount of gas than can be transformed into stars. Several authors (Madau, Ferrara \& Rees 2001; 
Ricotti \& Gnedin 2005; Okamoto, Gao \& Theuns 2008) agree that in these systems the SF 
efficiency decreases $\propto T_{vir}^3$. A suitable form is then
\be
\epsilon \propto \epsilon_* \big[1+(\frac{T_{vir}}{2\times 10^4{\rm K}})^3\big]^{-1},
\ee
where $\epsilon_*$ is the value determined by matching the global properties of the MW.
\section{Results}
The first test of the model is a comparison of the observed metallicity-luminosity relation with 
our predictions; the results are reported in Fig.~2. The overall agreement is satisfactory, and the
increasing trend of metallicity with luminosity well reproduced. The faint end of the relation, 
$L<10^6L_{\odot}$, is predominantly populated by minihaloes; above that luminosity H-cooling haloes
dominate. The low-$L$ segregation of minihaloes naturally arises from their low masses and it is further 
extended towards very faint luminosities, $L<10^4 L_{\odot}$, by the reduced SF efficiency and, most 
importantly, by their low initial baryonic fraction $f_b/f_c<0.1$. Being UFs defined as systems with 
$L<10^5 L_{\odot}$, we conclude that all UFs are minihaloes but there are minihaloes that are not UFs.

We have just pointed out that the initial baryonic fraction is a key factor in dSph galaxy evolution. 
What sets such initial value? Gas-rich systems are distributed through the entire mass 
($M_h=10^{7-10}\Msun$) and formation redshift ($z=2-10$) ranges of selected dSph candidates (Fig.~1). 
These dSphs are the present-day counterpart of predominantly newly virialized objects accreting gas from 
the GM. Intermediate systems ($0.1 < f_b/f_c < 0.5$) originate from mixed merging of star-forming and 
starless progenitors; they are typically more massive than gas-rich haloes and form at lower redshifts. 
Their smaller baryonic content is the result of shock-heating of the infalling gas during major merging 
events (Cox et~al. 2004) which stops accretion. Finally, gas-poor systems formed by merging of 
{\it recently} virialized, $M_h>M_{sf}$ progenitors. Since most of the diffuse gas is still accreting, 
shock heating quenches the infall early on. Note that a low baryonic content might also result from 
shock-stripping of the gas due to winds outflowing from nearby galaxies (Scannapieco, Ferrara \& 
Broadhurst 2000).

The faintest UFs are found to be gas-poor systems; this fact has two important implications. 
First, being the SN rate depressed in these systems by the scarce availability of gas, mechanical 
feedback has negligible effects, and they evolve as a closed-box. As essentially all metals are retained 
they have a relatively high Fe-abundance which is seen as an almost inverted Fe-L relation below 
$L=10^{4.5}L_{\odot}$. Second UFs have extremely large $M/L>100$ ratios, and the faintest ones among them 
reach such extreme values as $M/L\approx 10^4$.

To get more insight in the nature of UFs it is instructive to analyze their stellar MDF.
In Fig.~3 (left panel) we show the MDF averaged over all UF candidates in 10 realizations of the merger 
tree, along with the available data (Kirby et~al. 2008). The shape of the distribution is relatively 
broad extending from [Fe/H]$\approx -4$ to $\approx -1$, in good agreement with observations. This is 
noticeably different from the Sculptor MDF shown in the right panel of the same Figure. SN feedback in 
Sculptor is much stronger than in UFs, as explained above; SF activity is then terminated shortly after 
100~Myr (see Fig.~3 of SFS08) whereas UFs can continue to quietly form stars up to 1~Gyr from their birth. 
These physical differences are reflected in the wider UF MDF. In our model Sculptor-like 
dSphs\footnote{$L\approx 10^{6.2} L_{\odot}$, $\langle$[Fe/H]$\rangle\approx -1.8$, see also Fig.~2} are 
associated with gas-rich, H-cooling haloes, with $M_h \approx 10^8\Msun$, virializing at $z\approx 7.5$, 
in agreement with the findings in SFS08. However, in contrast with that study we now predict a small tail 
of [Fe/H]$<-3$ stars. These are relics of the rare SF episodes occurred in some progenitor minihaloes 
(which were not considered in SFS08) at $z>7.5$. Therefore, such a small number of very iron-poor 
stars in classical dSphs are expected to be characterized by the same abundance pattern of [Fe/H]$<-3$ 
stars in UFs. The rest of the MDF, on the contrary, is built after the stellar population bulk assembling 
at lower redshifts.
A striking difference between the two distributions in Fig.~3 is the larger fraction ($\approx 25\%$) of 
very metal poor stars ([Fe/H]$<-3$) present in the UF MDF. The interpretation of this result is clear: 
as UFs form earlier the metallicity of the GM gas out of which they virialize is correspondingly lower but 
still high enough that low-mass stars can be produced according to the critical metallicity criterion. 
In principle, then, UFs are potentially powerful benchmarks to validate this criterion, which would exclude 
the presence of stars of {\it total} metallicity below $Z_{cr}$, here fixed at $Z_{cr}=10^{-3.8}\Zsun$ 
(see SFS08 for details).
\section{Discussion}
\begin{figure}
  \centerline{\psfig{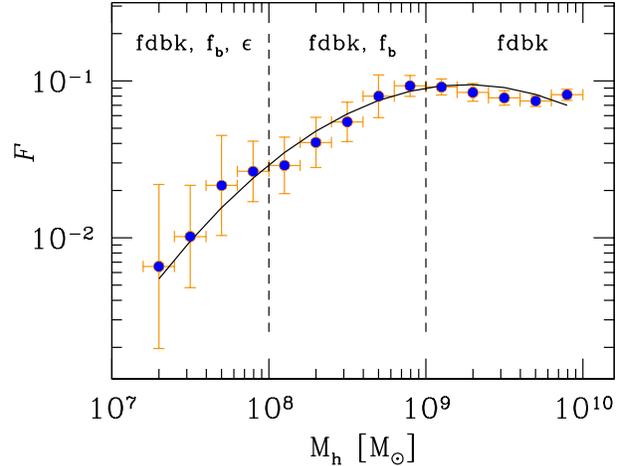}}
  \caption{Fraction ${\cal{F}} = M_*/f_c M_h$ of the potentially available cosmic baryon content turned 
    into stars as a function of $M_h$. The points are the average over all dSph candidates in 10 
    realizations of the merger tree. Error bars show the $\pm 1\sigma$ dispersion among different dSphs. 
    The solid line is the handy fit.} 
  \label{fig:4}
\end{figure}
UFs are the oldest and most dark matter ($M/L > 100$) dominated dSphs in the MW system with a total mass
$M= 10^{7-8}\Msun$ and $L=10^{3-5}L_{\odot}$; they are found to be left-overs of H$_2$-cooling minihaloes 
formed at $z > 8.5$, i.e. before reionization. Their MDF is broader (because of their more prolonged SF) 
and shifted towards lower [Fe/H] (as a result of a lower GM metallicity at the time of formation) with 
respect to classical dSphs.

The SF in the faintest UFs is depressed by two factors: the limited availability of cold, star-forming 
gas due to ineffective H$_2$ cooling, and their specific formation mechanism resulting in gas-starved 
systems. The mass of stars formed at $z=0$ normalized to the cosmic baryon content associated with a 
halo of mass $M_h$, ${\cal{F}} = M_*/f_c M_h$ is shown in Fig.~4 as a function of $M_h$. A handy fit to 
the curve is given by
\be
{\cal F} = exp(a+bx+cx^2) 
\ee
where $a=-66$, $b=5.97$, $c=-0.14$ and $x=ln(M_h/M_{\odot})$. By inspecting that Figure the interplay
between different factors affecting SF can be readily understood. Haloes above $10^9 \Msun$ convert about 
$10\%$ of their potentially available $f_c M_h$ baryonic mass into stars; this value is solely determined 
by mechanical feedback. In haloes with $10^{8}-10^9 \Msun$ gas infall starts to be quenched by 
shock-heating during formation by mergers, with SN feedback playing a sub-dominant role. Therefore 
${\cal F}$ drops by about a factor 3 in this range. Finally, minihaloes $M_h<10^8 \Msun$ are affected 
by an additional suppression factor related to their decreasing SF efficiency (eq.~1) due to radiative 
feedback acting on the H$_2$ chemical network. The larger ${\cal F}$ scatter towards low $M_h$ is hence 
due to the increasing number of physical processes influencing star formation.

The recent work of Madau et~al. (2008) attempts to determine ${\cal F}$ for minihaloes by matching the 
luminosity function of MW satellites in the SDSS under the assumption that the formation of these 
systems stopped at $z_{rei}=11$ due to reionization photoheating. They find that ${\cal F}=(0,0.0025,0.02)$ 
for haloes with a total mass $M_h=(<3.5\times 10^7\Msun,(3.5-7)\times 10^7\Msun,>7\times 10^7 \Msun)$. 
In the same range of masses we find slightly higher ${\cal F}$ values (Fig.~4); this difference is likely 
due to the assumption made here that $z_{rei}=6$. It is however unclear what is the effect of an early, 
$z>11$, formation on the median and shape of the minihaloes/UFs MDF; on the other hand our model cannot 
predict if the higher ${\cal F}$ produces an excess of visible MW satellites. Clearly a more comprehensive 
study is required to answer these questions. 

Our conclusions support the ``primordial scenario`` for the origin of dSphs proposed by Bovill \& Ricotti 
(2008). In such a scenario feedback effects of the type discussed here represent fundamental evolutionary 
ingredients. The picture is strengthen by the successful simultaneous match of the Fe-L relation and the 
MDF of UFs. Our model reproduces at the same time also the observed MDF of both the MW and of a 
prototypical classical dSph as Sculptor.

The skeptic might wonder about the possible role of tidal stripping. In our picture UFs are, among all 
dSphs, those associated with the highest$-\sigma$ fluctuations ($\sim 2\sigma$) forming at the earliest 
possible epochs (Fig.~1). According to N-body cosmological simulations (Diemand, Madau \& Moore 2005) 
this implies that UFs are most probably found at small Galactocentric radii, as indeed deduced from 
observations; such data however must be interpreted with care as it might be biased due to the 
magnitude limit of the SDSS. The proximity to the MW may cause gas/stellar loss by tidal stripping from 
these satellites (Mayer et~al. 2007). However, it seems unlikely that UF can be the stripped remnant of 
classical dSph as the scaling of the luminosity-velocity dispersion with luminosity would be too steep 
to explain the observed trend (Pe\~narrubia, Navarro \& McConnachie, 2008). The success of our model also 
lends support to the conclusion that tidal stripping plays at most a minor role.

A final caveat concerns the adopted heuristic assumption for $M_{sf}(z)$. Physically $M_{sf}(z)$ is 
tightly related to the reionization history of the MW environment. Although we have not attempted to 
model in detail the radiative feedback processes determining the evolution of such quantity, which we 
defer to further study, it is conceivable that the grow of the LW background intensity will suppress 
the cooling and SF ability of progressively more massive minihaloes. Guided by this general argument 
we have then chosen an heuristic form of $M_{sf}(z)$ which suitably accounts for this physical process
as required by the data. A more physical interpretation of its shape derived from a detailed modeling 
of the LW background intensity growth during reionization is nevertheless necessary; it may well unfold 
the complicated physics behind radiative feedback.

\section*{Acknowledgements}
We are grateful to A. Helmi, Y.~S. Li, L.~V. Sales, R. Schneider, E. Starkenburg, E. Tolstoy and to 
all DAVID\footnote{{\tt www.arcetri.astro.it/science/cosmology/index.html}} members for enlightening 
discussions. This work has been completed while the authors were visiting the Kapteyn Astronomical 
Institute whose hospitality is warmly acknowledged.

\bibliographystyle{mn}
\bibliography{biblio}

\label{lastpage}

\end{document}